\documentclass[a4paper]{article}
\usepackage{graphicx}
\usepackage{amssymb}
\usepackage{amsmath}
\usepackage{cite}
\usepackage[left=2cm,top=2cm,right=2cm,bottom=2cm,nohead]{geometry}
\usepackage[pdftex,unicode,pdfstartview=FitH,colorlinks=true,pdfborder={0 0 0 0},urlcolor=blue]{hyperref}
\usepackage[labelsep=space,format=plain,margin=0.5cm,textformat=simple,justification=justified,small,parskip=0cm]{caption}
\usepackage[onehalfspacing]{setspace}
\begin{document}
\title{Continuous transition from the extensive to the non-extensive statistics in an agent-based herding model}
\author{A. Kononovicius, J. Ruseckas}
\date{}

\maketitle

\begin{abstract}
Systems with long-range interactions often exhibit power-law distributions and can by described by the non-extensive statistical mechanics framework proposed by Tsallis. In this contribution we consider a simple model reproducing continuous transition from the extensive to the non-extensive statistics. The considered model is composed of agents interacting among themselves on a certain network topology. To generate the underlying network we propose a new network formation algorithm, in which the mean degree scales sub-linearly with a number of nodes in the network (the scaling depends on a single parameter). By changing this parameter we are able to continuously transition from short-range to long-range interactions in the agent-based model.
\end{abstract}

\section{Introduction}

Properties of systems with long-range interactions concern a wide
range of problems in physics \cite{Dauxois2002}: gravitational forces
\cite{Padmanabhan1990} and Coulomb forces in globally charged systems
\cite{Nicholson1983}, vortices in two-dimensional fluid mechanics
\cite{Chavanis2001}, wave-particles interaction \cite{Barre2004},
and trapped charged particles \cite{Elskens2002}. Such systems are
of particular interest because they violate extensivity and additivity,
two basic properties used to derive the thermodynamics of a system.
Consequently they have been a subject of extensive studies in the
recent years (for reviews see \cite{Campa2009,Bouchet2010}). Small
systems, in which the range of interactions is comparable to the size
of the system, are also non-additive and thus are similar to large
systems with truly long-range interactions. These systems can exhibit
novel types of behavior - e.g., inequivalence of the microcanonical
and canonical ensembles \cite{Mukamel2005} and negative microcanonical
specific heat \cite{Thirring1970}. Models with long-range interactions
often possess dynamical features like slow relaxation \cite{Mukamel2005,Dauxois2002}
and broken ergodicity \cite{Mukamel2005,Bouchet2008}. Another characteristic
feature is the emergence of long-lived non-equilibrium quasistationary
states (QSS) and violent relaxation into these states \cite{Lynden-Bell1967}.
Non-Gaussian distributions \cite{Latora2001} and non-exponential
relaxations for autocorrelations \cite{Yamaguchi2003} have been observed
as well.

Non-extensive statistical mechanics is intended to describe some
of the systems with long-range interactions by generalizing the Boltzmann-Gibbs
statistics \cite{Tsallis2009-1,Tsallis2009-2,Telesca2010}. There
are systems that, depending on the initial conditions, are not ergodic
in the entire phase space and may prefer a particular subspace. If
that subspace has a scale invariant geometry, a hierarchical or multifractal
structure, then the model points toward non-extensive statistical
mechanics. The generalized statistical mechanics framework is based
on a generalized entropy \cite{Tsallis2009-1}, which is assumed to
be given by
\begin{equation}
S_{q}={\textstyle \left(1-\int[p(x)]^{q}dx\right)/(q-1)}\,,\label{eq:q-entr}
\end{equation}
where $p(x)$ is a probability density function of finding the system
in the state characterized by the parameter $x$, while $q$ is a
parameter describing the non-extensiveness of the system. In the limit
$q\rightarrow1$ the traditional Boltzmann-Gibbs entropy is recovered
from Eq.~(\ref{eq:q-entr}) \cite{Tsallis2009-1,Tsallis2009-2}.
Concepts drawn from this generalized framework have found their applications
in a variety of traditional disciplines, such as physics \cite{Liu2008PhysRevLett,Pickup2009PhysRevLett,Beck2013PhysRevE},
chemistry, biology or economics, and also in an interdisciplinary
field of the complex systems \cite{Gell-Mann2004,Abe2006,Picoli2009}.

Consequences of long-range interactions usually have been investigated
in Hamiltonian systems. In this paper we explore long-range interactions
in agent-based modeling instead. Agent-based modeling is one the most
prominent contemporary tools used to obtain insights into the complex
socio-economic systems. It is the main tool used to model opinion
dynamics \cite{Borghesi2012PLOS,Vitanov2012Springer}, explain emergent
phenomena in microeconomics \cite{Yakovenko2009RMP} and macroeconomics
\cite{Tramontana2010RevEco,Westerhoff2010NJP}, reproduce the dynamics
observed in the financial markets \cite{Cristelli2012Fermi,Chakraborti2011RQUF2}
and solve logistic problems for the business practitioners \cite{Frederick2013PNAS}.
Some approaches starting from agent-based modeling obtain non-linear
stochastic differential equations (SDEs) as a macroscopic model for
the underlying agent-based dynamics \cite{Alfarano2008,Cristelli2012Fermi,Kononovicius2012PhysA,Feng2012PNAS},
thus providing microscoping reasoning for the socio-economic dynamics.
Another layer of understanding may be provided by another contemporary
tool known as network theory, which allows to uncover the intrinsic relationships
in geological \cite{Tenenbaum2012}, biological \cite{Bashan2012},
socio-economic \cite{Alfarano2009Dyncon,Biondo2013PhysRevE}
and other complex systems \cite{Barzel2013,Newman2010}.

In the context of this contribution the most interesting approaches are based on
the agent-based herding model, originally proposed and developed in a series of
papers by Kirman \cite{Kirman1991,Kirman1993,Kirman2002}, as these approaches
are able to reproduce both the power-law and Gaussian-like distributions
\cite{Alfarano2009Dyncon,Kononovicius2013SocTech}. In \cite{Alfarano2009Dyncon}
it was shown that Kirman's model reproduces power-law distribution if the
underlying model topology is a random network, but if the topology is a
small-world or a scale-free network, then the Gaussian-like distribution is
obtained. This result can be easily understood by looking into the scaling of
each network's mean degree. The network where the mean degree $\langle d\rangle$
is fixed, $\langle d\rangle\sim\mathrm{const}$, (e.g.,\ small-world or a
scale-free network) represents short-range interactions, whereas the network
where the mean degree scales linearly with the number of nodes, $\langle
d\rangle\sim N$, (e.g.,\ a random network) represents truly long-range
interactions and corresponds to Hamiltonian mean-field models.

In this paper we connect those two extreme cases by proposing a new network
formation model, which exhibits a sub-linear scaling of the mean degree,
$\langle d\rangle\sim N^{\alpha}$ (with $\alpha\in[0,1]$). By changing the
single network parameter $\alpha$ we can continuously transition from
short-range to long-range interactions in our agent-based model. This network
formation model connects random and scale-free networks and can be useful in
describing socio-economical systems.

The paper is organized as follows. In Section~\ref{sec:extensive} we describe an
extensive agent-based model corresponding to short-range interactions between
the agents. To investigate the transition to long-range interactions we consider
agent-based model implemented on a network. The network formation model is discussed
in Section~\ref{sec:network}. This model is able to produce hybrid networks in between
well-known random and scale-free networks and exhibits sub-linear scaling of the mean
degree with the increasing number of nodes. In Section~\ref{sec:agents} we investigate an
agent-based model implemented on this network. For this model the detailed
network structure is not important and mean-field approximation yields a good
result. Section \ref{sec:concl} summarizes our findings.

\section{Extensive agent-based model}

\label{sec:extensive}We consider an agent model similar to the model proposed by
Alan Kirman (see \cite{Kirman1993}). There is a fixed number of agents, $N$, each of them being in state $1$
or in state $2$. In this model dynamic evolution is described as a Markov chain,
the agents switch state either due to
idiosyncratic factors or under the influence (e.g., peer pressure) of other agents.
The lack of memory of the agents is the crucial
assumption to formalize the dynamics as a Markov process. Describing the
dynamics as a jump Markov process in a continuous time, we choose $\eta_1$ and $\eta_2$
to represent per-agent transition rates to the state written in the subscript. Namely, $\eta_1$
is a transition rate from state $2$ to state $1$. By choosing $n$ to represent a whole number of
agents in state $1$, it becomes convenient to obtain a number of agents in state $2$ via $N-n$.
The aforementioned transition rates $\eta_1$ and $\eta_2$ can depend on $n$, $N-n$ as well as
on the total number of agents $N$.

We can write the aggregate transition rates for one agent switching as
\begin{align}
p(n\rightarrow n+1) & \equiv p^{+}(n) = (N-n)\eta_1 \,,\\
p(n\rightarrow n-1) & \equiv p^{-}(n) = n \eta_2 \,.
\end{align}
The above probabilities define a one-step stochastic process
\cite{VanKampen2007NorthHolland}. The transition probabilities imply the Master
equation for the probability $P_{n}(t)$ to find $n$ agents in the state $1$ at
time $t$ \cite{VanKampen2007NorthHolland}:
\begin{equation}
\frac{\partial}{\partial t}P_{n}=p^{+}(n-1)P_{n-1}+p^{-}(n+1)P_{n+1} -(p^{+}(n)+p^{-}(n))P_{n}\,.\label{eq:master}
\end{equation}
For large enough $N$ we can represent the macroscopic system state by using a
continuous variable $x=n/N$. Using the birth-death process formalism
\cite{VanKampen2007NorthHolland}, one can obtain a non-linear Fokker-Planck
equation from the Master equation~(\ref{eq:master}) assuming that $N$ is large
and neglecting the terms of the Taylor expansion of the order of $1/N^{2}$:
\begin{equation}
\frac{\partial}{\partial t}P_{x}(x,t)=
\frac{\partial}{\partial x}[x\eta_{2}-(1-x)\eta_{1}]P_{x}(x,t)
+\frac{1}{2N}\frac{\partial^{2}}{\partial x^{2}}[(1-x)\eta_{1}+x\eta_{2}]P_{x}(x,t)\,.
\label{eq:FP-2}
\end{equation}
Taking into accound the diffusion term, the steady state solution of the
Fokker-Planck equation (\ref{eq:FP-2}) is
\begin{equation}
P_{0}(x)=\frac{C}{(1-x)\eta_{1}+x\eta_{2}}
\exp\left[-2N\int^{x}
\frac{x'\eta_{2}-(1-x')\eta_{1}}{(1-x')\eta_{1}+x'\eta_{2}}dx'\right]
\label{eq:steady-pdf}
\end{equation}

When the interactions between agents are short-range (in other words agents
interact in their fixed size local neighborhood), the model is extensive and the
transition rates $\eta_1$ and $\eta_2$ depend only on the continuous system
state variable $x=n/N$ and do not directly depend on total number of particles
$N$: $\eta_1=\eta_1(x)$ and $\eta_2=\eta_2(x)$. In the thermodynamic limit, when
$N\rightarrow\infty$, we can neglect the diffusion therm in Eq.~(\ref{eq:FP-2}).
In that case we get
\begin{equation}
\frac{\partial}{\partial t}P_{x}=\frac{\partial}{\partial x}[x\eta_{2}-(1-x)\eta_{1}]P_{x}
\end{equation}
with the corresponding steady state solution
\begin{equation}
P_{0}(x)=\delta(x-x_{0})
\end{equation}
Here $x_{0}$ is the solution of the equation describing the detailed balance:
\begin{equation}
x_{0}\eta_{2}(x_{0})=(1-x_{0})\eta_{1}(x_{0})\,.
\label{eq:det-balance}
\end{equation}

Taking into accound the diffusion term the steady state solution is given by
Eq.~(\ref{eq:steady-pdf}). When $x'=x_{0}$ then the expression in the integral
in Eq.~(\ref{eq:steady-pdf}) is zero. Expanding the expression in the integral
around the point $x'=x_{0}$ and keeping only first-order term we get
\begin{align}
P_{0}(x) \approx & C'\exp\left[-2N\int^{x}A(x'-x_{0})dx'\right] \nonumber\\
= &\sqrt{\frac{NA}{\pi}}\exp\left[-NA(x-x_{0})^{2}\right]
\end{align}
where the expansion coefficient $A$ is
\begin{equation}
A=g'(x_{0})+\frac{1}{2x_{0}(1-x_{0})}
\label{eq:a-extensive}
\end{equation}
with
\begin{equation}
e^{2g(x)}\equiv\frac{\eta_{2}(x)}{\eta_{1}(x)}
\end{equation}
We obtain that the steady state probability distribution function (PDF) is
approximately Gaussian with the width proportional to $1/\sqrt{N}$.
This result is in agreement with the research presented by Traulsen
\cite{Traulsen2005PhysRevLett,Traulsen2006PhysRevE,Traulsen2012PhysRevE},
who studied a very similar, yet significantly narrower (fixed form of $\eta_i$), case.

In order to investigate the effects of long-range interactions and
non-extensivity in the agent model we need to have the transition rates $\eta_1$
and $\eta_2$ that explicitly depend on the total number of agents $N$. To
construct the model that can have a practical relevance we will consider an
agent-based model implemented on a network. We start by proposing a new
network formation model in the following Section.

\section{Network formation model exhibiting sub-linear scaling of the mean degree}

\begin{figure}
\centering
\includegraphics[width=0.45\textwidth]{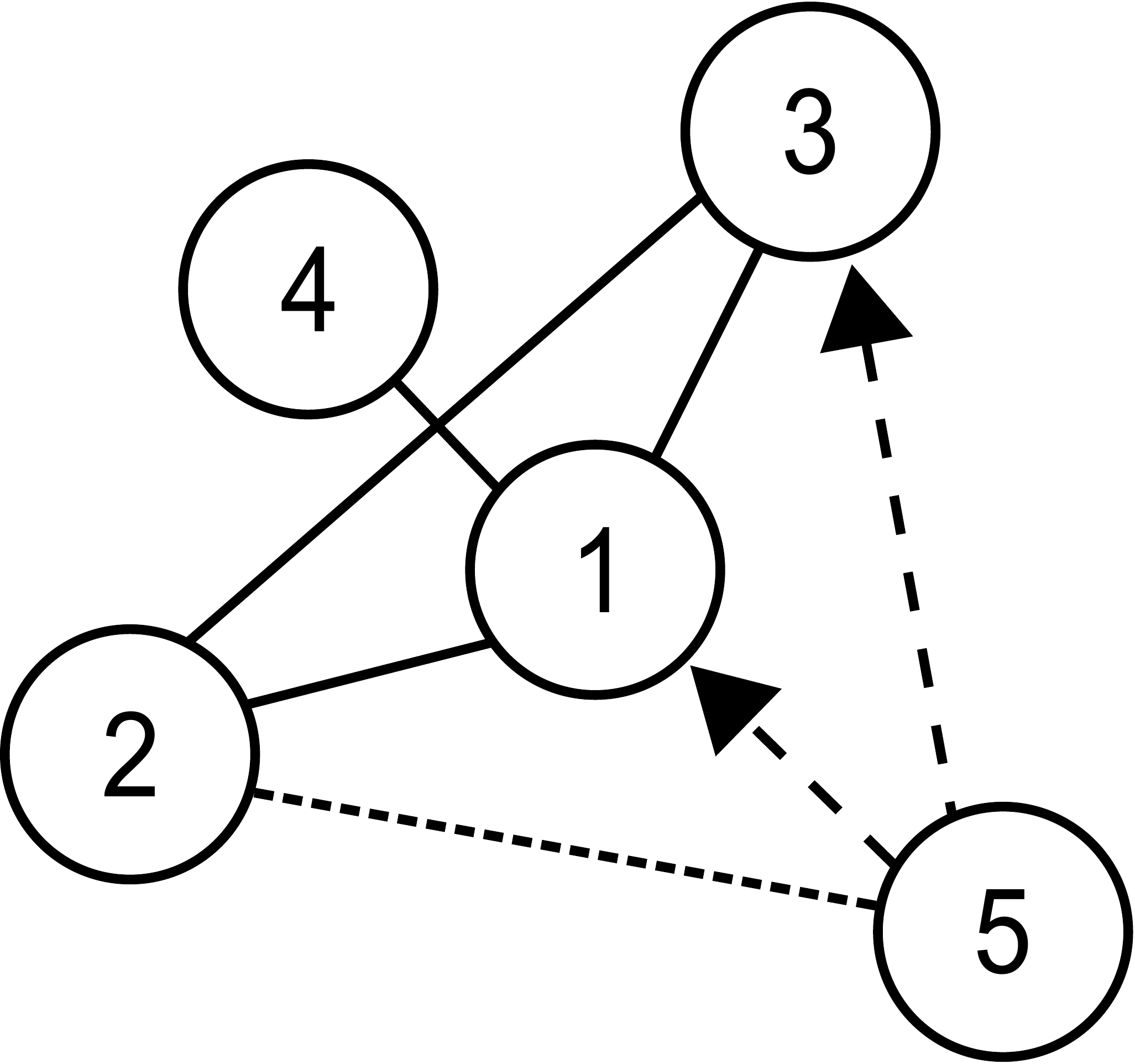}
\caption{Node~5 joins an existing network by making connection to Node~2 via
the ``rich gets richer'' scheme (dashed line without arrows). After
making this initial connection to the network, with a certain probability
given by Eq.~(\ref{eq:randomConnProb}) Node~5 may connect (dashed
arrows) to the neighbors of Node~2 (Node~1 and Node~3). Node~4 remains
intact as it is not a direct neighbor of Node~2.}
\label{fig:network-model}
\end{figure}

\label{sec:network} In this Section we propose a new network formation model
exhibiting sub-linear scaling of the mean degree $\langle d\rangle$ with the increasing
number of nodes $N$ in the network. To construct our network formation model we
have chosen the Barabasi-Albert model \cite{Albert2002RevModPhys} as our base
model. We extend this model by adding an additional step. This means that during
the first step in our network formation model
we add a new node to the network and connect it to one old node
based on the linear ``rich gets richer'' scheme. During the additional step the
new node may form additional links with the immediate neighbors of the old node,
the one it was connected to during the first step, each link is formed with probability
\begin{equation} p=p_{0}d^{-\gamma}\,,\label{eq:randomConnProb} \end{equation}
where $p_{0}$ is a probability to make a random connection when $\gamma=0$, $d$
is a degree of the old node, $\gamma$ is a probability scaling exponent, which
is related to the mean degree scaling exponent, $\alpha$. An exemplary schema of
the proposed formation model is shown in Fig.~\ref{fig:network-model}.

Note that the additional step is somewhat similar to the techniques used in the triad formation
\cite{Holme2002PhysRevE,Moriano2013JStatMech}, friends of friends
\cite{Jackson2007AER} and forest fire \cite{Leskovec2007TKDD} network formation
models. As in the works \cite{Holme2002PhysRevE,Moriano2013JStatMech,Jackson2007AER} the
additional links are formed only with the immediate neighbors of the old node.
Though unlike in \cite{Jackson2007AER} we use Barabasi-Albert model as a base
model. We also add a random amount of links during the additional step unlike the models
considered in
\cite{Holme2002PhysRevE,Moriano2013JStatMech,Jackson2007AER}. The forest fire
algorithm \cite{Leskovec2007TKDD} also adds a random number of links, but it
considers $\gamma=0$ case. In the forest fire algorithm the mean degree scaling
is achieved not by scaling the probability of forming the additional links, but
by repeating the additional step until no new links are formed. Note that there
a more network formation models, which exhibit sub-linear scaling of the mean
degree, but mostly they are overly general and lack connections to the actual
processes in the socio-economic systems
\cite{Akoglu2009DMKD,Bonato2010MSM,Colman2013PhysA}.

\begin{figure}
\centering
\includegraphics[width=0.45\textwidth]{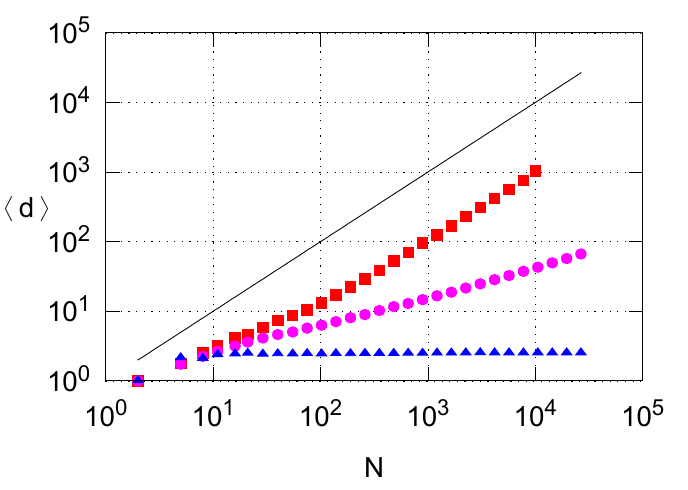}
\caption{Mean degree scaling for different values of $\gamma$: $0$ (red squares),
$0.3$ (magenta circles), $1$ (blue triangles). Black curve shows
the mean degree scaling in completely connected network.}
\label{fig:mean-deg}
\end{figure}

Dependence of the mean degree $\langle d\rangle$ on the number of nodes in the
network $N$ for various values of the parameter $\gamma$ is shown in
Fig.~\ref{fig:mean-deg}. Our numerical calculations indicate that
\begin{equation}
\alpha\approx(1-\gamma)^{2}
\end{equation}
for $\gamma\in[0,1]$.

\begin{figure}
\centering
\includegraphics[width=0.45\textwidth]{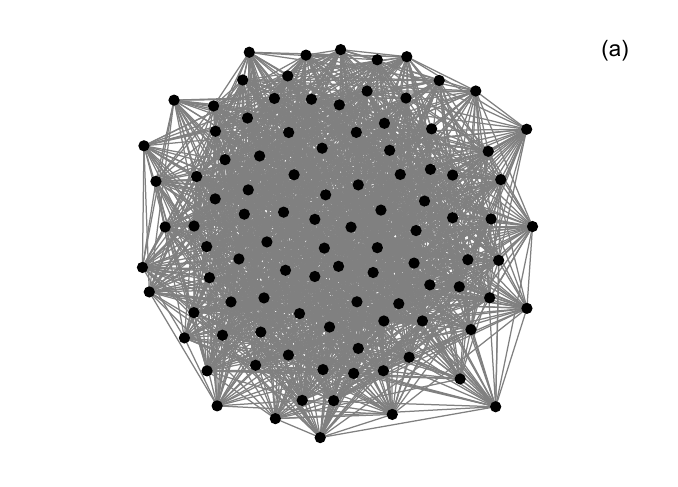}\includegraphics[width=0.45\textwidth]{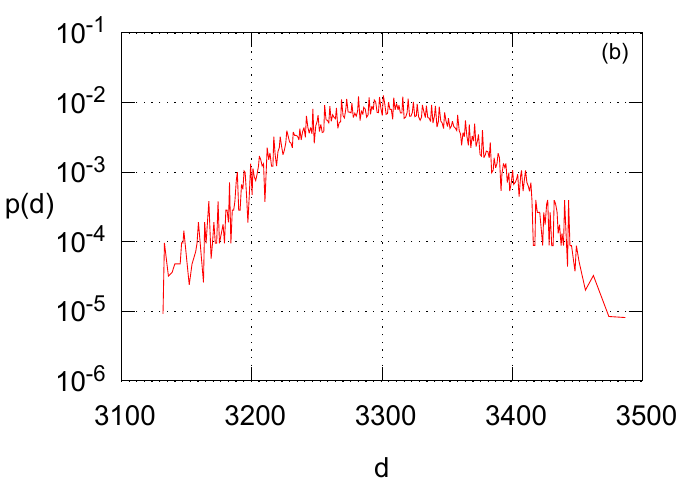}

\includegraphics[width=0.45\textwidth]{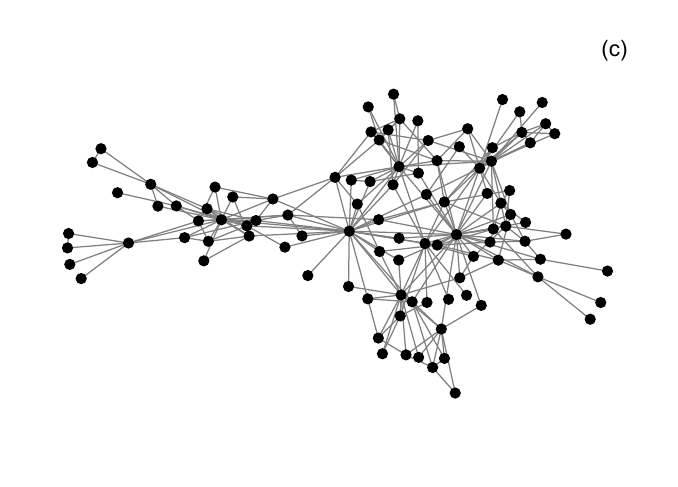}\includegraphics[width=0.45\textwidth]{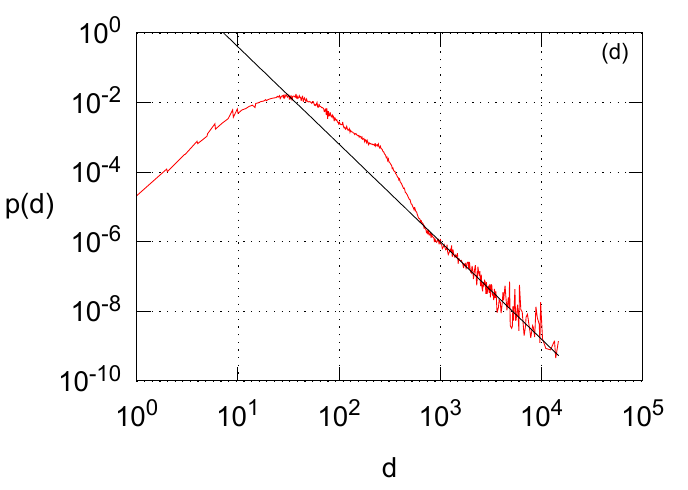}

\includegraphics[width=0.45\textwidth]{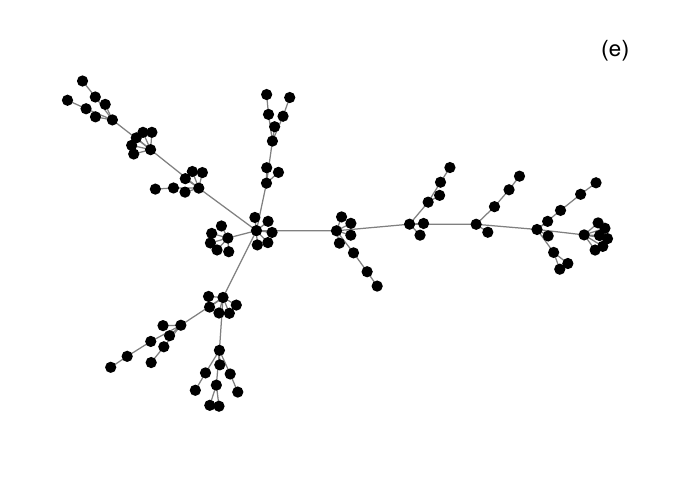}\includegraphics[width=0.45\textwidth]{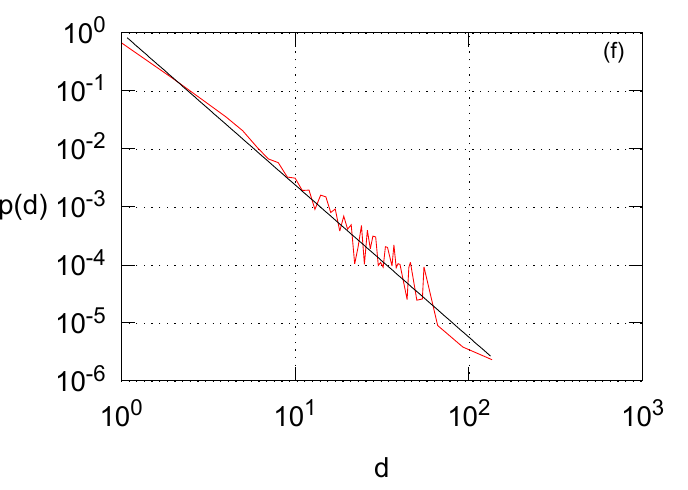}

\caption{Random network (topology (a), degree distribution (b)), scale-free
network (topology (e), degree distribution (f)) and hybrid network (topology
(c), degree distribution (d)) generated using the proposed network formation
algorithm. Network snapshots (a), (c), (e) were taken at $N=100$. Degree PDFs
were obtained on networks with $N=10^{4}$ (random network) and $N=3\cdot10^{4}$
(hybrid and scale-free networks). Black curves in (d) and (f) provide power law
fit (with exponent $\lambda=3$) for the tail of the degree PDF. Following
parameters were used: $p_{0}=0.3$, $\gamma=0$ (random network), $0.3$ (hybrid
network), $1$ (scale-free network).}
\label{fig:network-regime}
\end{figure}

In Fig.~\ref{fig:network-regime} we demonstrate a transition between
random and scale-free networks obtained using our network formation
algorithm. With small values of $\gamma$ we observe a randomly connected
clump of nodes, we also observe Gaussian-like degree distribution
in this clump. Slightly larger $\gamma$ values allow for large degree
hubs to form, while apparently random links are still present. While
from $\gamma\gtrsim 1$ the probability of forming random links becomes
small, thus random links disappear and only scale-free structure remains.
Mean degree scaling for the same values of $\gamma$ is shown in Fig.~\ref{fig:mean-deg}.

\section{Agent-based model executed on the network structure}

\label{sec:agents} In this Section we consider Kirman's agent-based model
implemented on a network generated using algorithm described in the previous Section.
There is a fixed number of agents, $N$, located on the
nodes of the network, each of them being in state $1$ or in state $2$. Note, as
the comparison of mean-field approximation with the exact solution shows, that
the detailed network structure for this model is not important. The main
influence of the network is via the scaling of the mean degree $\langle
d\rangle$ with the number of nodes $N$ in the network. Describing the dynamics
as a jump Markov process in a continuous time, the transition probabilities per
unit time for agent $i$ being in the state $X$ ($X=1,2$) to switch state to the
other state $Y$ ($Y\neq X$) are given by
\begin{equation}
p_{i}(X\rightarrow Y)= \sigma+h n_{i}(Y)\,,\label{eq:p-one}
\end{equation}
where $\sigma$ is the idiosyncratic switching rate, $h$ describes
the herding tendency and $n_{i}(Y)$ is the number of neighbors in
the state $Y$.

\subsection{Mean-field approximation}

The mean-field approach for the model yields the following
mean per-agent transition (from state $X$ to state $Y$) rates \cite{Alfarano2009Dyncon}:
\begin{equation}
\langle p_{i}(X\rightarrow Y)\rangle=\sigma+h\langle d\rangle \frac{N_{Y}}{N}\,,\label{eq:p-avg}
\end{equation}
where $N_{Y}$ is a total number of agents in the state $Y$. Using the notation introduced in
Section~\ref{sec:extensive} we would have $\eta_1 = \langle p_{i}(2\rightarrow 1)\rangle$ and 
$\eta_2 = \langle p_{i}(1\rightarrow 2)\rangle$.

Note that in the infinitely large system limit, $N\rightarrow\infty$,
the herding behavior term disappears if $\langle d\rangle\sim\mathrm{const}$,
while it remains constant if $\langle d\rangle\sim N$. If the herding
term disappears, or becomes negligible,
then the mean behavior of system becomes deterministic and only a
small Gaussian-like fluctuations occur (see Section~\ref{sec:extensive}), while otherwise the power-law
distribution is obtained \cite{Alfarano2009Dyncon,Kononovicius2013SocTech}.

The Fokker-Planck equation (\ref{eq:FP-2}) for the model now becomes
\begin{multline}
\frac{\partial}{\partial t}P_{x}(x,t)=-\frac{\partial}{\partial x}\sigma(1-2x)P_{x}(x,t)
+\frac{1}{2N}\frac{\partial^{2}}{\partial x^{2}}(2h\langle d\rangle x(1-x)+\sigma)
P_{x}(x,t)\,.
\label{eq:f-p-full}
\end{multline}
The dynamics of the continuous macroscopic system state variable $x$ can be modeled by the
SDE corresponding to the Fokker-Planck equation (\ref{eq:f-p-full}):
\begin{equation}
dx=\sigma(1-2x)dt+\sqrt{\frac{1}{N}(2h\langle d\rangle x(1-x)+\sigma)}dW_{t}\,,\label{eq:sde-herding}
\end{equation}
where $W_{t}$ is a Wiener process. In \cite{Ruseckas2011} it has been shown that
in the case when $\langle d\rangle\sim N$ the fluctuations of the ratio
$N_{2}/N_{1}$, exhibit $1/f^{\beta}$ power spectral density in a wide region of
frequencies growing with $N$. In particular, we have $1/f$ noise when
$\sigma/(d_{0}h)=2$. This is not the case when $\alpha<1$ because for $\alpha<1$
in the limit of $N\rightarrow\infty$ the macroscopic fluctuations of $x$ vanish.

\subsection{Steady state distribution of agents}

\begin{figure}
\centering
\includegraphics[width=0.45\textwidth]{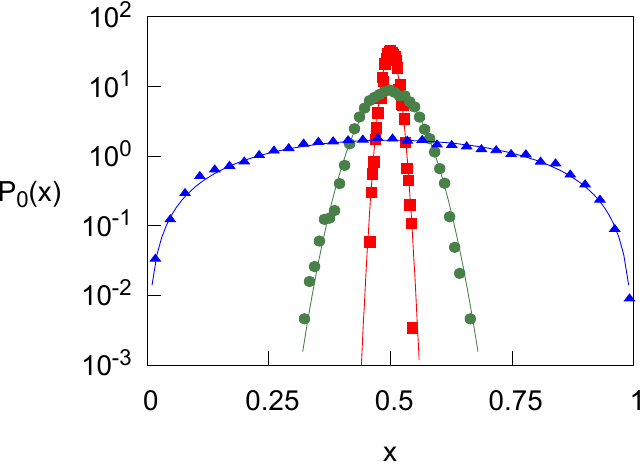}
\caption{Simulated steady state probability density function $P_{0}(x)$ for agent-based
model with different values of mean degree exponent $\alpha$: red
squares $\alpha=0$, green circles $\alpha=0.5$, blue triangles $\alpha=1$.
Solid lines show the mean-field approximation of the steady state
probability density function provided by Eq.~(\ref{eq:pdf}). The parameter values of the
model were $\sigma=1.5$, $h=1$, $N=3000$, $p_{0}=0.75$, $\Delta t=2\cdot10^{-5}$.
The parameter $\gamma$ values were $\gamma=1$ ($\alpha=0$), $\gamma=0.3$
($\alpha=0.5$), $\gamma=0$ ($\alpha=1$). From the scaling of the
mean degree $\langle d\rangle$ with changing $N$ the following $d_{0}$
values were obtained: $d_{0}=3.2$ ($\alpha=0$), $d_{0}=1.24$ ($\alpha=0.5$)
and $d_{0}=0.6$ ($\alpha=1$).}
\label{fig:difr-alpha}
\end{figure}

\begin{figure}
\centering
\includegraphics[width=0.45\textwidth]{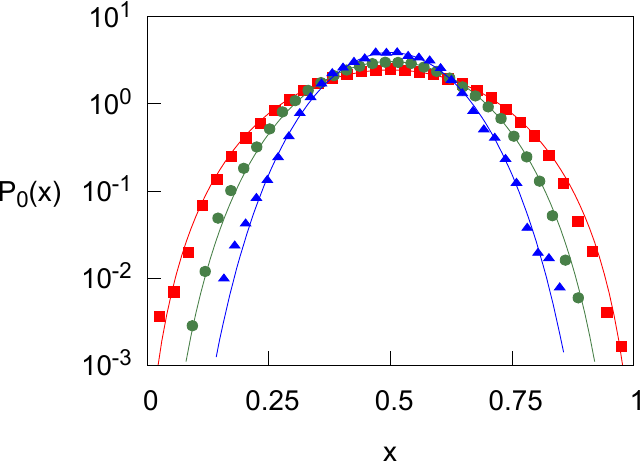}
\caption{Scaling of the simulated steady state probability density function $P_{0}(x)$
with the increasing number of agents in the model: red squares $N=100$,
green circles $N=500$, blue triangles $N=3000$. Solid lines show
mean-field approximation of the steady state probability density function provided by Eq.~(\ref{eq:pdf}).
The remaining parameters of the model were $\sigma=1.5$, $h=1$,
$p_{0}=0.75$, $\gamma=0.15$, $\Delta t=2\cdot10^{-5}$. The value
of $d_{0}=0.9$ was obtained from the scaling of the mean degree $\langle d\rangle$
with changing $N$.}
\label{fig:difr-n}
\end{figure}

Now let us consider the steady state of this system of agents and
investigate the probability density function $P_{0}(x)$. If
the mean degree $\langle d\rangle$ scales as $N^{\alpha}$, that
is $\langle d\rangle=d_{0}N^{\alpha}$, the steady state PDF obtained
from Eq.~(\ref{eq:f-p-full}) according to Eq.~(\ref{eq:steady-pdf}) is
\begin{equation}
P_{0}(x)=C[\varepsilon+2N^{\alpha}x(1-x)]^{\varepsilon N^{1-\alpha}-1}\,,\label{eq:pdf}
\end{equation}
where
\begin{equation}
\varepsilon\equiv\frac{\sigma}{d_{0}h}
\end{equation}
and $C$ is the normalization constant. The steady state PDF obtained from
numerical simulation of the agent-based model described by Eq.~(\ref{eq:p-one})
and comparison with the mean-field approximation (\ref{eq:pdf}) is shown in
Figs.~\ref{fig:difr-alpha} and \ref{fig:difr-n}. In the numerical simulations we
choose a fixed time step $\Delta t$ and consider transition probabilities equal
to $p_{i}(X\rightarrow Y)\Delta t$. The time step must be chosen such that all
transition probabilities should be between $0$ and $1$. For a given network
structure, we synchronously update the state of each agent according to the
transition probabilities. In the mean-field steady state PDF we use the
parameter $d_{0}$ extracted from the scaling of the mean degree $\langle
d\rangle$ of the network with the number of nodes $N$. We see a good agreement
of the simulated PDF with the mean-field approximation. The width of the steady
state PDF increases with increase of $\alpha$, as is shown in
Fig.~\ref{fig:difr-alpha} and decreases with increase of the number of agents
$N$, as is evident from Fig.~\ref{fig:difr-n}. In the limit of
$N\rightarrow\infty$ the PDF $P_{0}(x)$ becomes very narrow if $\alpha<1$.

Eq.~(\ref{eq:pdf}) can be rewritten in a $q$-Gaussian form
\begin{equation}
P_{0}(x)=C^{\prime}\exp_{q}\left[-A_{q}\left(x-\frac{1}{2}\right)^{2}\right]\label{eq:pdf-2}
\end{equation}
with
\begin{equation}
q=1-\frac{1}{\varepsilon N^{1-\alpha}-1}\,,\qquad A_{q}=2N^{1-\alpha}\frac{1-\frac{1}{\varepsilon}N^{\alpha-1}}{\frac{1}{2\varepsilon}+N^{-\alpha}}\,.\label{eq:pdf-q}
\end{equation}
The $q$-Gaussian PDF can be obtained by applying the standard variational
principle on the generalized entropy (\ref{eq:q-entr}) (see \cite{Tsallis2009-1}).
In the above $\exp_{q}(\cdot)$ is the $q$-exponential function,
defined as 
\begin{equation}
\exp_{q}(x)\equiv[1+(1-q)x]_{+}^{\frac{1}{1-q}}\,,\label{eq:q-exp1}
\end{equation}
here $[x]_{+}=x$ if $x>0$, and $[x]_{+}=0$ otherwise. The $q$-Gaussian
steady state solution of the Fokker-Planck equation (\ref{eq:f-p-full})
can be explained by noting that Eq.~(\ref{eq:f-p-full}) satisfies
the condition given by Eq.~(11) of Ref.~\cite{Borland1998a}. The
steady state PDF (\ref{eq:pdf-2}) having $q$-Gaussian form for finite
values of $N$ is in agreement with known results that Tsallis generalized
canonical distribution describes systems in contact with a finite
heath bath \cite{Plastino1994,Potiguar2003}. Eq.~(\ref{eq:pdf-2})
also confirms the similarity of small systems to large systems with
truly long-range interactions.

If the interactions are long-range, $\alpha=1$ and $\langle d\rangle\sim N$,
and the system is infinitely large, $N\rightarrow\infty$, then the
steady-state PDF (\ref{eq:pdf}) has a power-law form 
\begin{equation}
P_{0}(x)=\frac{\Gamma(2\varepsilon)}{\Gamma(\varepsilon)^{2}}[x(1-x)]^{\varepsilon-1}\,.\label{eq:steady-a1}
\end{equation}
This corresponds to non-extensivity parameter
\begin{equation}
q=1-\frac{1}{\varepsilon-1}\,.
\end{equation}
On the other hand, if interactions are short-range, $\alpha=0$ and $\langle
d\rangle\sim\mathrm{const}$, and the system infinitely large,
$N\rightarrow\infty$, then according to Eq.~(\ref{eq:det-balance}), the
steady-state PDF is Dirac delta function centered on $x_{0}=1/2$. As real
systems are never infinite, for large $N$ the steady-state PDF has a
Gaussian-like form. If $\alpha<1$ and $N$ is large then $q$ tends to $1$ and
from the properties of the $q$-exponential function we get that the steady state
PDF (\ref{eq:pdf-2}) is approximately Gaussian
\begin{equation}
P_{0}(x)\sim\exp\left[-N^{1-\alpha}A\left(x-\frac{1}{2}\right)^{2}\right]
\end{equation}
with
\begin{equation}
A=\begin{cases}
\frac{2}{\frac{1}{2\varepsilon}+1}\,, & \alpha=0\\
4\varepsilon\,, & 0<\alpha<1
\end{cases}\label{eq:aa}
\end{equation}
In this equation the coefficient $A$ for $\alpha=0$ is the same as given by
Eq.~(\ref{eq:a-extensive}). Thus the steady-state PDF retains its form in the
$N\rightarrow\infty$ limit only if $\alpha=1$, while in all other cases the
$N$-dependece problem, consider by Alfarano and Milakovic
\cite{Alfarano2009Dyncon}, is obtained: namely the shape and variance of the
distribution is lost with the increasing size of the system. It should be noted,
that when $0<\alpha<1$, the fluctuations in the system decay not as
$1/\sqrt{N}$, as it is usual in the statistics of extensive systems, but slower
as $1/\sqrt{N^{1-\alpha}}$. The fluctuations decay slower with increasing $N$
when $\alpha$ is closer to $1$. In the limiting non-extensive case of $\alpha=1$
the fluctuations do not decay at all with increasing the system size and are
always macroscopic.

\section{Conclusions}

\begin{figure}
\centering
\includegraphics[width=0.45\textwidth]{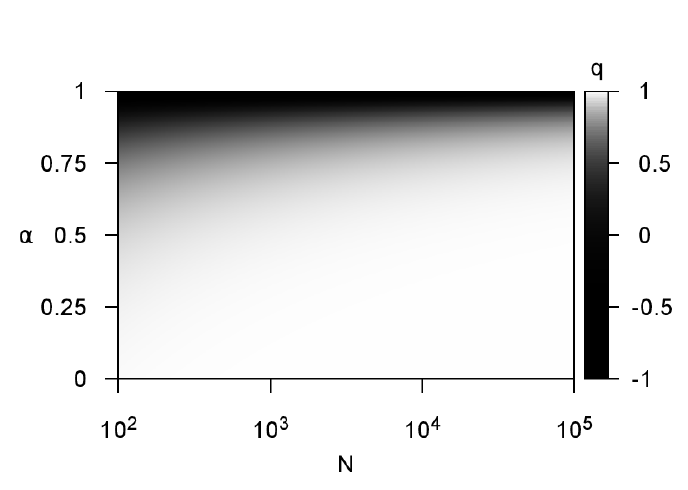}
\caption{Regions of extensive and non-extensive behavior of the agent-based
model as described by the non-extensivity parameter $q$, Eq.~(\ref{eq:pdf-q}).
White color corresponds to extensive ($q\approx1$), black color to
non-extensive ($q<1$) behavior. The parameter $\varepsilon$ equals
$1.5$.}
\label{fig:ext-next}
\end{figure}

\label{sec:concl} In summary, we have demonstrated a simple agent-based model
that by changing the single parameter $\alpha$ can continuously transition from
extensive to non-extensive statistics. Transition from extensive to
non-extensive statistics in the agent-based model with changing of the parameter
$\alpha$ and the number of agents $N$ is shown in Fig.~\ref{fig:ext-next}. As we
can see, the extensive region becomes wider as $N$ increases. However, for
$\alpha=1$ the behavior is non-extensive for all values of $N$.

The steady state distribution of agents for a finite system size is described by
$q$-Gaussian (\ref{eq:pdf-2}) with $q\leq1$. For $\alpha<1$ and increasingly
large system size (e.q. $N\rightarrow\infty$) the steady state distribution of
the model tends to a Gaussian form with the width depending on $\alpha$: as
$\alpha$ increases the width decreases more slowly with increasing $N$. This
simple model allows us to deepen the understanding of the effects of long-range
interactions and observe the emergence of non-extensivity.


\begin{thebibliography}{10}
\providecommand{\url}[1]{#1}
\csname url@samestyle\endcsname
\providecommand{\newblock}{\relax}
\providecommand{\bibinfo}[2]{#2}
\providecommand{\BIBentrySTDinterwordspacing}{\spaceskip=0pt\relax}
\providecommand{\BIBentryALTinterwordstretchfactor}{4}
\providecommand{\BIBentryALTinterwordspacing}{\spaceskip=\fontdimen2\font plus
\BIBentryALTinterwordstretchfactor\fontdimen3\font minus
  \fontdimen4\font\relax}
\providecommand{\BIBforeignlanguage}[2]{{%
\expandafter\ifx\csname l@#1\endcsname\relax
\typeout{** WARNING: IEEEtran.bst: No hyphenation pattern has been}%
\typeout{** loaded for the language `#1'. Using the pattern for}%
\typeout{** the default language instead.}%
\else
\language=\csname l@#1\endcsname
\fi
#2}}
\providecommand{\BIBdecl}{\relax}
\BIBdecl

\bibitem{Dauxois2002}
T.~Dauxois, S.~Ruffo, E.~Arimondo, and M.~Wilkens, Eds., \emph{Dynamics and
  Thermodynamics of Systems with Long-Range Interactions}, ser. Lect. Not.
  Phys.\hskip 1em plus 0.5em minus 0.4em\relax New York: Springer, 2002, vol.
  602.

\bibitem{Padmanabhan1990}
T.~Padmanabhan, \emph{Phys. Rep.}, vol. 188, p. 285, 1990.

\bibitem{Nicholson1983}
D.~R. Nicholson, \emph{Introduction to Plasma Theory}.\hskip 1em plus 0.5em
  minus 0.4em\relax New York: John Wiley, 1983.

\bibitem{Chavanis2001}
P.~H. Chavanis and C.~Sire, \emph{Phys. Fluids}, vol.~13, p. 71804, 2001.

\bibitem{Barre2004}
J.~Barr{\'e}, T.~Dauxois, G.~{De Ninno}, D.~Fanelli, and S.~Ruffo, \emph{Phys.
  Rev. E}, vol.~69, p. 045501(R), 2004.

\bibitem{Elskens2002}
Y.~Elskens and D.~F. Escande, \emph{Microscopic Dynamics of Plasmas and
  Chaos}.\hskip 1em plus 0.5em minus 0.4em\relax Bristol: IOP, 2002.

\bibitem{Campa2009}
A.~Campa, T.~Dauxois, and S.~Ruffo, \emph{Phys. Rep.}, vol. 480, p.~57, 2009.

\bibitem{Bouchet2010}
F.~Bouchet, S.~Gupta, and D.~Mukamel, \emph{Physica A}, vol. 389, pp.
  4389--4405, 2010.

\bibitem{Mukamel2005}
D.~Mukamel, S.~Ruffo, and N.~Schreiber, \emph{Phys. Rev. Lett.}, vol.~95, p.
  240604, 2005.

\bibitem{Thirring1970}
W.~Thirring, \emph{Z. Phys.}, vol. 235, p. 339, 1970.

\bibitem{Bouchet2008}
F.~Bouchet, T.~Dauxois, D.~Mukamel, and S.~Ruffo, \emph{Phys. Rev. E}, vol.~77,
  p. 011125, 2008.

\bibitem{Lynden-Bell1967}
D.~Lynden-Bell, \emph{Mon. Not. R. Astron. Soc.}, vol. 136, p. 101, 1967.

\bibitem{Latora2001}
V.~Latora, A.~Rapisarda, and C.~Tsallis, \emph{Phys. Rev. E}, vol.~64, p.
  056134, 2001.

\bibitem{Yamaguchi2003}
Y.~Y. Yamaguchi, \emph{Phys. Rev. E}, vol.~68, p. 066210, 2003.

\bibitem{Tsallis2009-1}
C.~Tsallis, \emph{Introduction to Nonextensive Statistical Mechanics --
  Approaching a Complex World}.\hskip 1em plus 0.5em minus 0.4em\relax New
  York: Springer, 2009.

\bibitem{Tsallis2009-2}
------, \emph{Braz. J. Phys.}, vol.~39, p. 337, 2009.

\bibitem{Telesca2010}
L.~Telesca, \emph{Tectonophysics}, vol. 494, p. 155, 2010.

\bibitem{Liu2008PhysRevLett}
B.~Liu and J.~Goree, ``Superdiffusion and non-gaussian statistics in a
  driven-dissipative 2d dusty plasma,'' \emph{Phys. Rev. Lett.}, vol. 100, p.
  055003, 2008.

\bibitem{Pickup2009PhysRevLett}
R.~M. Pickup, R.~Cywinski, C.~Pappas, B.~Farago, and P.~Fouquet, ``Generalized
  spin-glass relaxation,'' \emph{Phys. Rev. Lett.}, vol. 102, p. 097202, 2009.

\bibitem{Beck2013PhysRevE}
C.~Beck and S.~Miah, ``Statistics of lagrangian quantum turbulence,''
  \emph{Phys. Rev. E}, vol.~87, p. 031002, 2013.

\bibitem{Gell-Mann2004}
C.~M. Gell-Mann and C.~Tsallis, \emph{Nonextensive Entropy---Interdisciplinary
  Applications}.\hskip 1em plus 0.5em minus 0.4em\relax NY: Oxford University
  Press, 2004.

\bibitem{Abe2006}
S.~Abe, \emph{Astrophys. Space Sci.}, vol. 305, p. 241, 2006.

\bibitem{Picoli2009}
S.~Picoli, R.~S. Mendes, L.~C. Malacarne, and R.~P.~B. Santos, \emph{Braz. J.
  Phys.}, vol.~39, p. 468, 2009.

\bibitem{Borghesi2012PLOS}
C.~Borghesi, J.-C. Raynal, and J.~P. Bouchaud, ``Election turnout statistics in
  many countries: similarities, differences, and a diffusive field model for
  decision-making,'' \emph{PLoS ONE}, vol.~7, p. e36289, 2012.

\bibitem{Vitanov2012Springer}
N.~K. Vitanov and M.~Ausloos, ``Knowledge epidemics and population dynamics
  models for describing idea diffusion,'' in \emph{Models of Science Dynamics},
  ser. Understanding Complex Systems, A.~Scharnhorst, K.~B{\"o"}rner, and
  P.~van~den Besselaar, Eds.\hskip 1em plus 0.5em minus 0.4em\relax Berlin,
  Germany: Springer, 2012, pp. 69--127.

\bibitem{Yakovenko2009RMP}
V.~M. Yakovenko and J.~B. Rosser, ``Colloquium: Statistical mechanics of money,
  wealth, and income,'' \emph{Reviews of Modern Physics}, vol.~81, p. 1703,
  2009.

\bibitem{Tramontana2010RevEco}
F.~Tramontana, ``Economics as a compartmental system: a simple macroeconomic
  example,'' \emph{International Review of Economics}, vol.~57, no.~4, pp.
  347--360, 2010.

\bibitem{Westerhoff2010NJP}
F.~Westerhoff, ``An agent-based macroeconomic model with interacting firms,
  socio-economic opinion formation and optimistic/pessimistic sales
  expectations,'' \emph{New Journal of Physics}, vol.~12, no.~7, p. 075035,
  2010.

\bibitem{Cristelli2012Fermi}
M.~Cristelli, L.~Pietronero, and A.~Zaccaria, ``Critical overview of
  agent-based models for economics,'' in \emph{Proceedings of the School of
  Physics {"}E. Fermi{"}, Course CLXXVI}, F.~Mallnace and H.~E. Stanley,
  Eds.\hskip 1em plus 0.5em minus 0.4em\relax Bologna-Amsterdam: SIF-IOS, 2012,
  pp. 235 -- 282.

\bibitem{Chakraborti2011RQUF2}
A.~Chakraborti, I.~M. Toke, M.~Patriarca, and F.~Abergel, ``Econophysics
  review: Ii. agent-based models,'' \emph{Quantitative Finance}, vol.~7, pp.
  1013--1041, 2011.

\bibitem{Frederick2013PNAS}
R.~Frederick, ``Agents of influence,'' \emph{Proceedings of the National
  Academy of Sciences of the United States of America}, vol. 110, no.~10, pp.
  3703--3705, 2013.

\bibitem{Alfarano2008}
S.~Alfarano, T.~Lux, and F.~Wagner, ``Time variation of higher moments in a
  financial market with heterogeneous agents: An analytical approach,''
  \emph{J. Econ. Dyn. Control}, vol.~32, p. 101, 2008.

\bibitem{Kononovicius2012PhysA}
A.~Kononovicius and V.~Gontis, ``Agent based reasoning for the non-linear
  stochastic models of long-range memory,'' \emph{Physica A}, vol. 391, pp.
  1309--1314, 2012.

\bibitem{Feng2012PNAS}
L.~Feng, B.~Li, B.~Podobnik, T.~Preis, and H.~E. Stanley, ``Linking agent-based
  models and stochastic models of financial markets,'' \emph{PNAS}, vol.~22,
  no. 109, pp. 8388--8393, 2012.

\bibitem{Tenenbaum2012}
J.~N. Tenenbaum, S.~Havlin, and H.~E. Stanley, ``Earthquake networks based on
  similar activity patterns,'' \emph{PhysRevE}, vol.~86, p. 046107, 2012.

\bibitem{Bashan2012}
A.~Bashan, R.~P. Bartsch, J.~W. Kantelhardt, S.~Havlin, and P.~C. Ivanov,
  ``Network physiology reveals relations between network topology and
  physiological function,'' \emph{Nature Communications}, vol.~3, p. 702, 2012.

\bibitem{Alfarano2009Dyncon}
S.~Alfarano and M.~Milakovic, ``Network structure and n-dependence in
  agent-based herding models,'' \emph{J. Econ. Dyn. Control}, vol.~33, pp.
  78--92, 2009.

\bibitem{Biondo2013PhysRevE}
A.~E. Biondo, A.~Pluchino, A.~Rapisarda, and D.~Helbing, ``Stopping financial
  avalanches by random trading,'' \emph{Phys. Rev. E}, vol.~88, p. 062814,
  2013.

\bibitem{Barzel2013}
B.~Barzel and A.~L. Barabasi, ``Universality in network dynamics,''
  \emph{Nature Physics}, vol.~9, p. 673–681, 2013.

\bibitem{Newman2010}
M.~E.~J. Newman, \emph{Networks: An Introduction}.\hskip 1em plus 0.5em minus
  0.4em\relax Oxford: Oxford University Press, 2010.

\bibitem{Kirman1991}
A.~Kirman, ``Epidemics of opinion and speculative bubbles in financial
  markets,'' in \emph{Money and Financial Markets}, M.~P. Taylor, Ed.\hskip 1em
  plus 0.5em minus 0.4em\relax Cambridge: Blackwell, 1991, p. 354.

\bibitem{Kirman1993}
------, ``Ants, rationality, and recruitment,'' \emph{Q. J. Econ.}, vol. 108,
  p. 137, 1993.

\bibitem{Kirman2002}
A.~Kirman and G.~Teyssi{\'e}re, ``Microeconomic models for long memory in the
  volatility of financial time series,'' \emph{Stud. Nonlinear Dyn. Econ.},
  vol.~5, p. 137, 2002.

\bibitem{Kononovicius2013SocTech}
A.~Kononovicius and V.~Daniunas, ``Agent-based and macroscopic modeling of the
  complex socio-economic systems,'' \emph{Soc. Tech.}, vol.~3, pp. 85--103,
  2013.

\bibitem{VanKampen2007NorthHolland}
N.~G. van Kampen, \emph{Stochastic process in Physics and Chemistry}.\hskip 1em
  plus 0.5em minus 0.4em\relax Amsterdam: North Holland, 2007.

\bibitem{Traulsen2005PhysRevLett}
A.~Traulsen, J.~C. Claussen, and C.~Hauert, ``Coevolutionary dynamics: From
  finite to infinite populations,'' \emph{Phys. Rev. Lett.}, vol.~95, p.
  238701, 2005.

\bibitem{Traulsen2006PhysRevE}
------, ``Coevolutionary dynamics in large, but finite populations,''
  \emph{Phys. Rev. E}, vol.~74, p. 011901, 2006.

\bibitem{Traulsen2012PhysRevE}
------, ``Stochastic differential equations for evolutionary dynamics with
  demographic noise and mutations,'' \emph{Phys. Rev. E}, vol.~85, p. 041901,
  2012.

\bibitem{Albert2002RevModPhys}
R.~Albert and A.~L. Barabasi, ``Statistical mechanics of complex networks,''
  \emph{Rev. Mod. Phys.}, vol.~74, pp. 47--97, 2002.

\bibitem{Holme2002PhysRevE}
P.~Holme and B.~J. Kim, ``Growing scale-free networks with tunable
  clustering,'' \emph{Phys. Rev. E}, vol.~65, no.~2, p. 026107, 2002.

\bibitem{Moriano2013JStatMech}
P.~Moriano and J.~Finke, ``On the formation of structure in growing networks,''
  \emph{J. Stat. Mech.}, vol. 2013, no.~6, p. P06010, 2013.

\bibitem{Jackson2007AER}
M.~O. Jackson and B.~W. Rogers, ``Meeting strangers and friends of friends: How
  random are social networks?'' \emph{Am. Econ. Rev.}, vol.~97, no.~3, pp.
  890--915, 2007.

\bibitem{Leskovec2007TKDD}
J.~Leskovec, J.~Kleinberg, and C.~Faloutsos, ``Graph evolution: Densification
  and shrinking diameters,'' \emph{TKDD}, vol.~1, no.~1, p. 1217301, 2007.

\bibitem{Akoglu2009DMKD}
L.~Akoglu and C.~Faloutsos, ``Rtg: a recursive realistic graph generator using
  random typing,'' \emph{Data Mining Knowledge Discovery}, vol.~19, no.~2, pp.
  194--209, 2009.

\bibitem{Bonato2010MSM}
A.~Bonato, J.~Janssen, and P.~Pralat, ``A geometric model for on-line social
  networks,'' in \emph{Proc. Int. Workshop on Modeling Social Media}.\hskip 1em
  plus 0.5em minus 0.4em\relax New York, NY, USA: ACM, 2010, p. 1835984.

\bibitem{Colman2013PhysA}
E.~R. Colman and G.~J. Rodgers, ``Complex scale-free networks with tunable
  power-law exponent and clustering,'' \emph{Physica A}, vol. 392, pp.
  5501--5510, 2013.

\bibitem{Ruseckas2011}
J.~Ruseckas, B.~Kaulakys, and V.~Gontis, \emph{EPL}, vol.~96, p. 60007, 2011.

\bibitem{Borland1998a}
L.~Borland, \emph{Phys. Lett. A}, vol. 245, pp. 67--72, 1998.

\bibitem{Plastino1994}
A.~R. Plastino and A.~Plastino, \emph{Phys. Lett. A}, vol. 193, p. 140, 1994.

\bibitem{Potiguar2003}
F.~Q. Potiguar and U.~M.~S. Costa, \emph{Physica A}, vol. 321, p. 482, 2003.

\end{thebibliography}
\end{document}